\documentclass[preprint,onecolumn,nofootinbib,superscriptaddress]{revtex4-1}

\usepackage{lineno}
\usepackage{braket}
\usepackage{mathtools}
\usepackage{ulem}
\usepackage{upgreek}

\newcommand{\nn}{\nonumber}

\newcommand{\be}{\begin{equation}}
\newcommand{\ee}{\end{equation}}

\usepackage{color}
\usepackage{subcaption}
\usepackage{caption}
\usepackage[font=small,labelfont=bf,justification=justified]{caption}
\definecolor{purple}{rgb}{.36,.12,.60}
\definecolor{orange}{rgb}{.9,.3,.0}

\begin{document}

\title {
Dynamics of Transverse Spin and Longitudinal Fields of Cylindrical Vector Beams in Optically Active Media}

\author{Yuanyang~Xie}

\author{Alexey~Krasavin}

\author{Anatoly~V.~Zayats}

\affiliation{Department of Physics and London Centre for Nanotechnology, King's College London, Strand, London WC2R 2LS, UK}

\author{Andrei Afanasev}

\affiliation{Department of Physics,
The George Washington University, Washington, DC 20052, USA}

%\affiliation{And ... ?}

\begin{abstract}
Due to the inhomogeneous polarisation across the beam profile, cylindrical vector beams interact with optically active media in a complex manner. Here, we analyse evolution of polarisation of cylindrical vector beams propagating in an isotropic optically-active medium. After identifying polarisation normal modes of three-dimensional electromagnetic fields, we predict periodic inter-conversion between azimuthally- and radially-polarised modes of the beams accompanied by rotation of the transverse optical spin and pulsing   field during the propagation. Theory and simulations are validated by experimental observations. The observed effects maybe important for imaging in biological chiral media, enhanced chiral sensing and enantioselective spectroscopy, nonlinear optics in chiral media, and generally enhanced spin-orbit coupling and nanoscale vector field engineering.   
%A longitudinal electric field shows oscillatory behavior, while the transverse spin undergoes rotation and reaches a maximum degree of polarisation (C-line) within a radial position from 0 to wavelength/$\pi$ during the oscillation period independently of beam's expansion due to diffraction. 

%which reveal propagation-independent magnitude ratio with respect to their transverse counterparts. 
\end{abstract}
\date{\today
}
\maketitle

%%%%%%%%%%%%%%%%%%%%%%%%%%%%%%%%%%
%{\it Notes

%Novelty and objectives: 

%- Dichroism of longitudinal field; 
%- oscillations in plane wave-- just phase difference between lcp and rcp, in the confined beams: the influence of the front curvature  and final size, to focus on the presence of the longitudinal component; 
%- Normal modes - azimuthal is not normal mode 

%- generation of longitudinal field in optically active (or chiral media) ? -- In the title ?

%- rotation of transverse spin in optically active medium as in neutron scattering through weak interactions .... 

%\
%}

\section{Introduction}

%Particle-like properties of light are  essential for new developments in photonics and quantum optics. The present study is focused on the particle-like behavior of optical spin during propagation of light in optically-active matter.
%evolution of cylindrical polarisation modes and 3D optical spin during propagation of structured light in matter.

%The history of the problem goes back to the foundations of electromagnetic theory. In 1845, Michael Faraday discovered \cite{Faraday1846} that magnetization of materials results in rotation of propagating light's polarisation plane, thereby relating magnetism and optics. In 1851, George G. Stokes introduced three parameters to describe light's polarisation \cite{stokes1851composition}, defining -- using a modern language -- a spin-density matrix of a photon. Later on, in 1861-62 \cite{Maxwell1861}, James~C.~Maxwell, at the time a Professor of Natural Philosophy in King's College London, concluded that Faraday's rotation is due to different response of the magnetized materials to left- and right-circularly polarised light, helping him to formulate a unified theory of electricity and magnetism.  

Photon spin plays an important role in modern optics. It can couple to chiral matter or through magneto-optical effects to magnetic media
%, thereby enabling control of light-matter interactions 
and provide control over directionality of guided modes in photonic circuits \cite{lodahl2017chiral,aiello2022chirality}. On the other hand, optical beams may be structured to carry spatially-distributed and topologically-protected spin information that can enhance the bandwidth and robustness of optical communications and further enable photon-based quantum technologies \cite{shen2024optical,lei2025topological}. 
Optical processes involving evanescent waves \cite{eismann2021transverse}, surface polaritons \cite{bliokh2012transverse}, phase-structured waves that carry an orbital angular momentum (OAM) \cite{bliokh2015transverse}, or, generally occurring at the nanoscale \cite{novotny2012principles}, can produce complex optical polarisation topologies, which cannot be described by conventional three Stokes parameters, and a full three-dimensional treatment of electromagnetic fields is required \cite{carozzi2000,setala2002degree, aiello2015transverse,Dennis_2004,varshalovich1988quantum, Afanasev_2020}. This includes three parameters of vector polarisation (or spin) and five parameters of polarisation tensor (or alignment, that describes how a material changes the polarisation state of light). Three degrees of freedom for the photonic spin include both longitudinal (as in a plane wave) and transverse (absent in a plane wave as it requires a longitudinal field) components with respect to the propagation direction and allow for formation of topological spin structures that can be associated with optical spin skyrmions and merons \cite{shi2021spin,cheng2025navigating,mata2025skyrmionic,annenkova2025universal}. 

% An important question on whether structured electromagnetic waves would lead to new effects in probing chirality of materials has been actively discussed in the literature \cite{forbes2019spin, capolino2024}.

%In view of high interest to applications of the three-dimensional properties of photon spin, a question of its interaction with matter becomes important. 
%Would the topological spin information be robust against propagation? Does it have particle-like properties in the interaction with matter? 

Cylindrical Vector Beams (CVBs) is an example of optical beams for which a three-dimensional nature of the electromagnetic field is essential \cite{Zhan:09}. 
The two archetypal examples of CVBs are the radially polarised (RP) and the azimuthally polarised (AP) beams. The former requires the longitudinal electric field in the centre of the beam while the latter supports longitudinal magnetic field. Using birefringent or dichroic plates, conventional RP-AP mode conversion can be achieved \cite{Fatemi:11}). Although propagation of vector beams and evolution of their polarisation were studied in atomic gases under application of an external magnetic field \cite{Stern:16,wang2020vectorial} or polarised optical pumping \cite{Wang:18}, their propagation in an optically active media allowing to address a fundamental question of evolution of transverse optical spin in the presence of optical activity has remained unexplored.

Here, we theoretically analysed and experimentally demonstrated polarisation properties of CVBs propagating in an isotropic medium with natural optical activity. Particularly, by identifying polarisation normal modes of 
%We also identified the normal modes of optically active media that have propagation-invariant polarisation structures. Using these normal modes of 
three-dimensional electromagnetic fields, we reveal and experimentally demonstrate the inter-conversion of radially-polarised and azimuthally-polarised modes of the vector beam, accompanied by the evolution of transverse optical spin and longitudinal electric field. We emphasise similarities between photon spin rotation under optical activity and neutron spin rotation induced by parity-violating weak interactions \cite{forte1980first,sushkov1982parity,klein1983neutron}. 

%Inter-modal conversion of AP and RP cylindrical beams using birefringent or dichroic plates is a known technique \cite{Fatemi:11}. Propagation of vector beams and evolution of their polarisation was observed in magnetised Rb gas \cite{Stern:16,wang2020vectorial} and in a pump-probe measurements of the atom gas \cite{Wang:18}. However, the dynamics of transverse optical spin of vectorial vortices in matter remained unexplored so far, serving as motivation for our study. 
%Particle-like properties of light are  essential for new developments in photonics and quantum optics. 

%he paper is organized as follows: Sec.~II presents a theoretical formalism, Sec.~III follows with simulations using COMSOL package, Sec.~IV describes experimental observations, and Sec.~V presents discussion and conclusions.

\section{Results}

Let us consider a lossless, isotropic optically-active dielectric medium and identify polarisation normal modes of non-planar electromagnetic waves. Normal modes have the polarisation state which remains unchanged during propagation.  
%The solution for the plane waves is known: polarisation state of left- and right circular polarisation (LHC and RHC). The approach is as follows: (a)~recognize (and verify) that circularly polarised Bessel beams are normal modes for propagation in optically active media, (b)~expand azimuthal and radial modes in terms of LHC and RHC Bessel solutions, and (c)~analyse the resulting evolution of polarisation under propagation.
They are the eigenstates of the curl operator $\vec\nabla\times$ and, therefore, diagonalise the constitutive relations in a medium (see Appendix A):
\begin{equation}
\label{eq:nmode}
 \vec\nabla\times\vec E=\pm k \vec E \, , 
\end{equation}
where $k$ is the wave vector. We consider this relation as a criterion for a normal mode. For the plane waves in optically active medium, normal modes are the fields with left- (LCP, $\sigma= 1$) and right- (RCP, $\sigma= -1$) circular polarisation. 

\subsection{Inter-modal oscillations in non-diffractive Bessel CVBs}

Non-diffractive CVBs can be represented by Bessel beams which are the exact solutions of a non-paraxial wave equation in cylindrical coordinates:
\begin{align}
\label{eq:twistedwf}
 \vec{E}_{\kappa m_\gamma k_z \sigma}(\vec r,t) &= -i \sigma \, E_0 \,
	e^{i(k_z z - \omega t + m_\gamma \varphi )}					\times \Bigg\{	e^{-i \sigma \varphi}  \cos^2\frac{\theta_k}{2} 	\,
	J_{m_\gamma-\sigma}(\kappa\rho) \, \vec{e}_\sigma 
				\nn\\
	& \qquad + \frac{i}{\sqrt{2}}  \sin\theta_k	\,
	J_{m_\gamma}(\kappa\rho) \, \vec{e}_0  
	-   e^{ i \sigma \varphi}  \sin^2\frac{\theta_k}{2} 	\,
	J_{m_\gamma+\sigma}(\kappa\rho) \, \vec{e}_{-\sigma}
	\Bigg\}	\,,
\end{align}
where $ \vec r=(\rho \cos\varphi,\rho\sin\varphi, z)$, $\kappa=k\sin \theta_k,\ k_z=k \cos\theta_k, \
\vec e_0 =  \hat{\vec z}, \text{and} \ \vec e_\sigma=-\frac{1}{\sqrt{2}}(\sigma \hat{\vec \rho} +i \hat{\vec \varphi})e^{i\sigma\varphi}$ define cylindrical basis, $E_0$ is the normalisation constant, $\omega$ is the angular frequency, and $J_n(\kappa\rho)$ are cylindrical Bessel functions of order $n$. Bessel beams in turn can be represented by a superposition of plane waves aligned on a conical surface with a polar angle $\theta_k$ with respect to the propagation axis $z$ \cite{Durnin87,Jaregui05,jentschura2011generation}. A Bessel beam (Eq.~\ref{eq:twistedwf}) that carries a projection of the total angular momentum $m_\gamma$ along the $z$-axis, formed by plane waves of a given helicity $\sigma=\pm1$, is a polarisation normal mode which satisfies the same normal-mode condition $\vec \nabla\times \vec E(\vec r)=\sigma k \vec{E}(\vec r)$ as individual plane waves from which it is formed (see also \cite{hanifeh2020optimally}).

Azimuthally and radially polarised Bessel beams can be formed by linear superpositions of the normal modes with $\sigma=\pm 1$ by setting $m_\gamma=0$ and fixing initial conditions  for the field amplitudes at $z=0$ and $t=0$: 
\begin{equation}
\label{eq:APRP0}
  \vec E_{AP}(0)=\frac{i}{\sqrt{2}}[ \vec E_+(0)+\vec E_-(0)] \, ,\ \ \
 \vec E_{RP}(0)=\frac{1}{\sqrt{2}}[-\vec E_+(0)+\vec E_-(0) ] \, ,  
\end{equation}
where we introduced a short-hand notation $\vec E_\pm(0)\equiv \vec E|_{(m_\gamma=0,\sigma=\pm 1,z=0,t=0)}$. 
%It follows from Eq.~(\ref{eq:twistedwf}) that  the vector $\vec E_{AP}(0)$ has only a $\varphi$-component, while $\vec E_{RP}(0)$ has both radial ($\rho$) and longitudinal ($z$) components.
Assigning wave vectors $k_+$ and $k_-$ to the normal modes $\vec E_+(z,t)$ and $\vec E_-(z,t)$, respectively, we find that the AP and RP modes undergo inter-modal conversion upon propagation by a distance $z$, therefore, are not normal modes of an optically active medium:
\begin{align}
\label{eq:APev}
 \vec E_{AP}(z,t)
 %&=  \frac{i}{\sqrt{2}}\left( \vec E_+(0) e^{i(k_+z-\omega t)}+\vec E_-(0) e^{i(k_-z-\omega t)}\right) \\ \nonumber
 %&=\frac{i}{\sqrt{2}}e^{i(\bar k z-\omega t)}\left( \vec E_+(0) e^{-iz\delta k}+\vec E_-(0) e^{iz\delta k}\right) \\ \nonumber
 &=e^{i(\bar k z-\omega t)}\left( \vec E_{AP}(0) \cos(z\delta k)-\vec E_{RP}(0)\sin(z\delta k)\right),\\ \nonumber
%\end{align}
% \begin{align}
%\label{eq:RPev}
 \vec E_{RP}(z,t)
 %&=  \frac{1}{\sqrt{2}}\left( -\vec E_+(0) e^{i(k_+z-\omega t)}+\vec E_-(0) e^{i(k_-z-\omega t)}\right) \\ \nonumber
 %&=\frac{1}{\sqrt{2}}e^{i(\bar k z-\omega t)}\left( -\vec E_+(0) e^{-iz\delta k}+\vec E_-(0) e^{iz\delta k}\right) \\ \nonumber
 &=e^{i(\bar k z-\omega t)}\left( \vec E_{RP}(0) \cos(z\delta k)+\vec E_{AP}(0)\sin(z\delta k)\right),
\end{align}
where $\bar k=(k_-+k_+)/2$ and $\delta k=(k_--k_+)/2$ is the {\it rotary power for plane waves} \cite{saleh2008fundamentals}.
%The relation to the {\it rotary power}  \cite{saleh2008fundamentals} is:
%\begin{equation}
%\label{eq:rotpower}
%    \rho_{rot}=\frac{\pi}{\lambda_0}(n_--n_+)\equiv \frac{1}{2}(k_--k_+)=\delta k,
%\end{equation}
%where $\lambda_0$ is wavelength in vacuum.
Equation (4) predicts oscillations between propagating AP and RP modes with an oscillation length given by $z_\text{osc}=\pi/\delta k$ (Fig.~\ref{fig:3DRot}, Fig.~\ref{fig:edens}a,b). The same length $z_\text{osc}$ is required for 180$^{\circ}$ rotation of the polarisation plane of a plane wave in the medium with the same rotatory power, 
%$c.f.$ \cite{saleh2008fundamentals}, 
relating these two different polarisation phenomena to a single dynamical origin -- optical activity of matter.
Polarisation normal modes of CVBs in an optically active medium were identified in Refs.~\cite{hanifeh2020optimally,ye2021enhancing} in the context of possible enhancement of light sensitivity to probing chirality of matter. (In Ref.~\cite{hanifeh2020optimally}, these modes were called ``optimally chiral".)
%because they satisfy a duality condition $E^\pm=\pm i\eta_0H^\pm$ between electric and magnetic fields and hence maximise chirality density of light. 

\subsection{Optical spin dynamics}
\begin{figure}[th]
     \begin{subfigure}[b]{0.46\textwidth}
        \includegraphics[width=\textwidth]{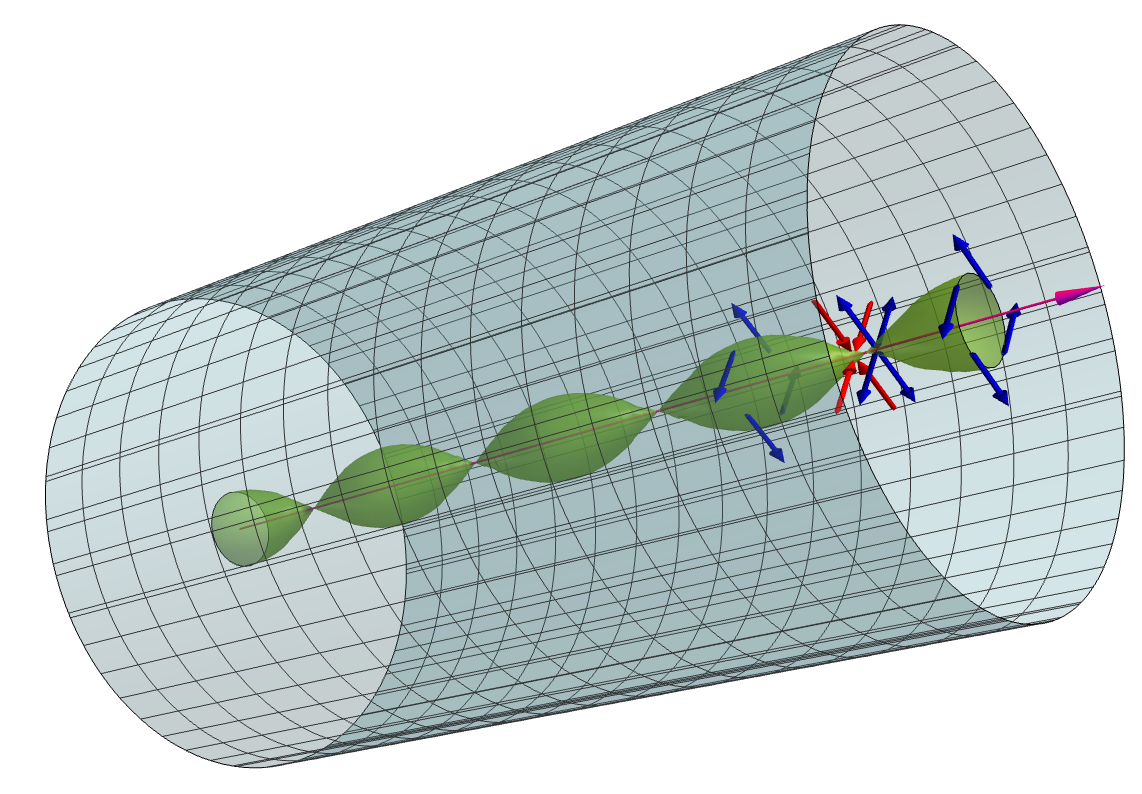}
        \caption{}
        \label{fig:a}
    \end{subfigure}
    \hfill
    \begin{subfigure}[b]{0.4\textwidth}
        \includegraphics[width=\textwidth]{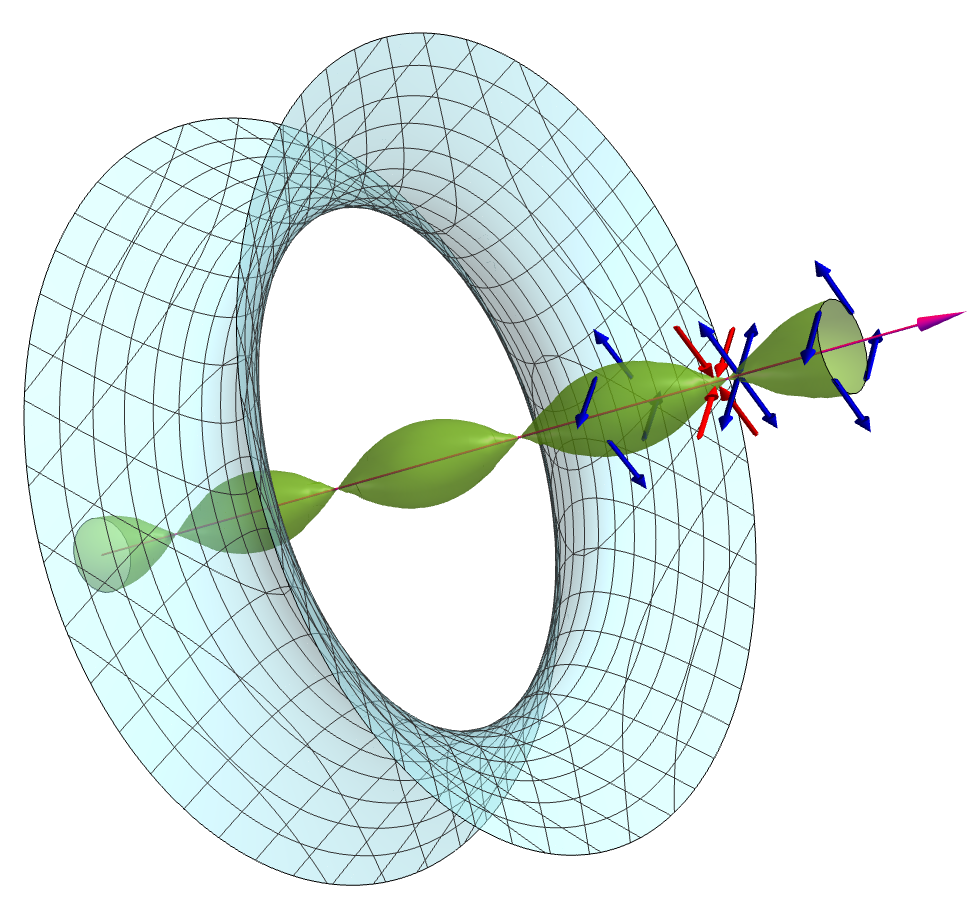}
        \caption{}
        \label{fig:b}
    \end{subfigure}
    \caption{\raggedright  Schematics of the evolution of (a) Bessel and (b) LG CVB parameters upon propagation in optical active medium.
    Blue surface represents a beam intensity profile that remains constant for Bessel and expands due to diffraction for LG beams. The C-surface ($S=1$), shown in green, remains almost identical for Bessel and LG beams, independently of the beam expansion; the C-surface is ``pulsating", with a radius shrinking to zero for pure AP states and reaching a maximum radius of $\lambda/\pi$ for pure RP states. Spin direction, shown by arrows, is azimuthal for the RP mode, becoming radial when approaching the AP state and flipping sign after passing through the AP state. Blue (red) arrows indicate a positive (negative) radial direction of spin.}
    \label{fig:3DRot}
\end{figure}

To better understand the dynamics of optical spin during the azimuthal-radial mode oscillations which inevitably involve evolution of the longitudinal electric field as it is present in RP modes and absent in AP modes, it is useful to consider the region near the beam centre in a paraxial approximation. Setting  $m_\gamma=0$ and  performing a Taylor series expansion of Eq.~(\ref{eq:twistedwf}) in the limit $\theta_k\ll 1$ while keeping $\rho=\text{const}$, we obtain in the leading order in $\theta_k$:
\begin{align}
\label{eq:normmode}
 \vec{E}_{\kappa m_\gamma k_z \sigma}(\vec r,t)|_{m_\gamma=0,\ \theta_k\ll 1} &=i E_0 \theta_k \left[\vec e_\sigma e^{-i\sigma\varphi}\frac{k\rho}{2}-\frac{i\sigma}{\sqrt{2}}\vec e_0\right]e^{i (k z-\omega t)} \\ \nonumber
 &=-\frac{i}{\sqrt{2}} E_0 \theta_k \left[\frac{k\rho}{2}(\sigma\hat{\vec \rho}+i\hat{\vec \varphi})+i\sigma\hat{\vec z}\right]e^{i (k z-\omega t)} .
\end{align}
This approximate expression describes the field in the region much smaller than the radius of a first intensity peak of the Bessel beam and explicitly satisfies both the normal mode condition (Eq.~\ref{eq:nmode}) and Gauss's law $\vec \nabla \vec E=0$, both of which would be violated without the longitudinal field component (described by $\vec e_0$ term) present. Transverse field components carry a common factor of $k\rho/2$ equal for the radial and azimuthal fields, but the longitudinal field is independent of $\rho$, making the relative strengths of the longitudinal and transverse field components linearly dependent on proximity to the beam axis.
Corresponding expressions for the AP and RP fields at $z=0$ and $t=0$ are
%\begin{align}
%    \hat{\vec \rho}&=\frac{1}{\sqrt{2}}(-\vec e_+e^{-i\varphi}+\vec e_- e^{i\varphi})\\ \nonumber
%    \hat{\vec \varphi}&=\frac{i}{\sqrt{2}}(\vec e_+e^{-i\varphi}+\vec e_- e^{i\varphi}),
%\end{align}
%we obtain at $z=0$, $t=0$:
 \begin{equation}
 \label{eq:EAPpar}
  \vec E_{AP}(0)|_{\theta_k\ll 1}=iE_0\theta_k \frac{\bar k\rho}{2}\hat{\vec\varphi}  
\end{equation}   
and %$i.e.$ the azimuthally-polarised field, and 
\begin{equation}
\label{eq:ERPpar}
  \vec E_{RP}(0)|_{\theta_k\ll 1}=iE_0\theta_k \left( \frac{\bar k\rho}{2}\hat{\vec\rho}+i\hat{\vec z}\right),  
\end{equation}
respectively; note an additional longitudinal field component for the radially-polarised mode. As follows from Eq.~(\ref{eq:APev}), the azimuthal and radial field components in the transverse plane remain in-phase during propagation, whereas the $z$-component of the field arising from the RP mode has its phase shifted by $\pi/2$. In the vicinity of the beam centre, where the presence of the longitudinal field is essential, this results in the formation of spin angular momentum in the transverse plane which spatial profile is periodic in $z$ with a period of $z_\text{osc}$, as described by Eq.~(\ref{eq:APev}) and shown in Fig.~\ref{fig:3DRot}.
%Applying Eqs.(\ref{eq:APev}), we can obtain simplified formulae for evolution of beam's polarisation parameters under medium's optical activity.

Dynamics of a three-dimensional electromagnetic field structure in the central region of the beam can be described by a ($3\times3$) spin density matrix formalism \cite{Afanasev_2020,varshalovich1988quantum,afanasev2023nondiffractive}. Expressing a polarisation coherence matrix in terms of state multipoles, we analyse normalised (electric) spin density $\vec S$ and an alignment parameter $p_{zz}$ that quantifies relative magnitudes of transverse $\vec E_\perp$ and longitudinal $E_z$ electric fields:
\begin{equation}
\label{eq:pzzDef}
    \vec S= \frac{i \vec E \times \vec E^*}{|\vec E|^2}, \ \ \ p_{zz}=\frac{|E_\perp|^2-2|E_z|^2}{|E_\perp|^2+|E_z|^2}.
\end{equation}

\begin{figure}[htbp]
    \centering
    \begin{subfigure}[b]{0.45\textwidth}
        \includegraphics[width=\textwidth]{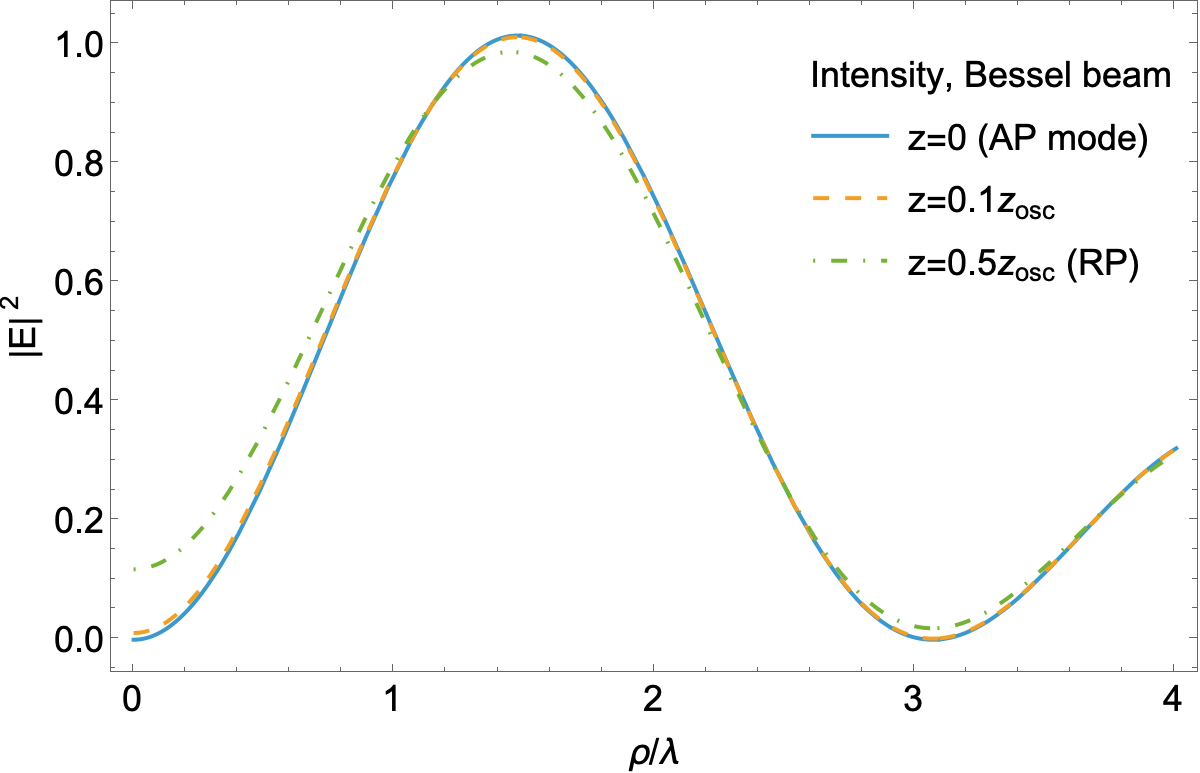}
        \caption{}
    %    \label{fig:a}
    \end{subfigure}
    \hfill
    \begin{subfigure}[b]{0.45\textwidth}
        \includegraphics[width=\textwidth]{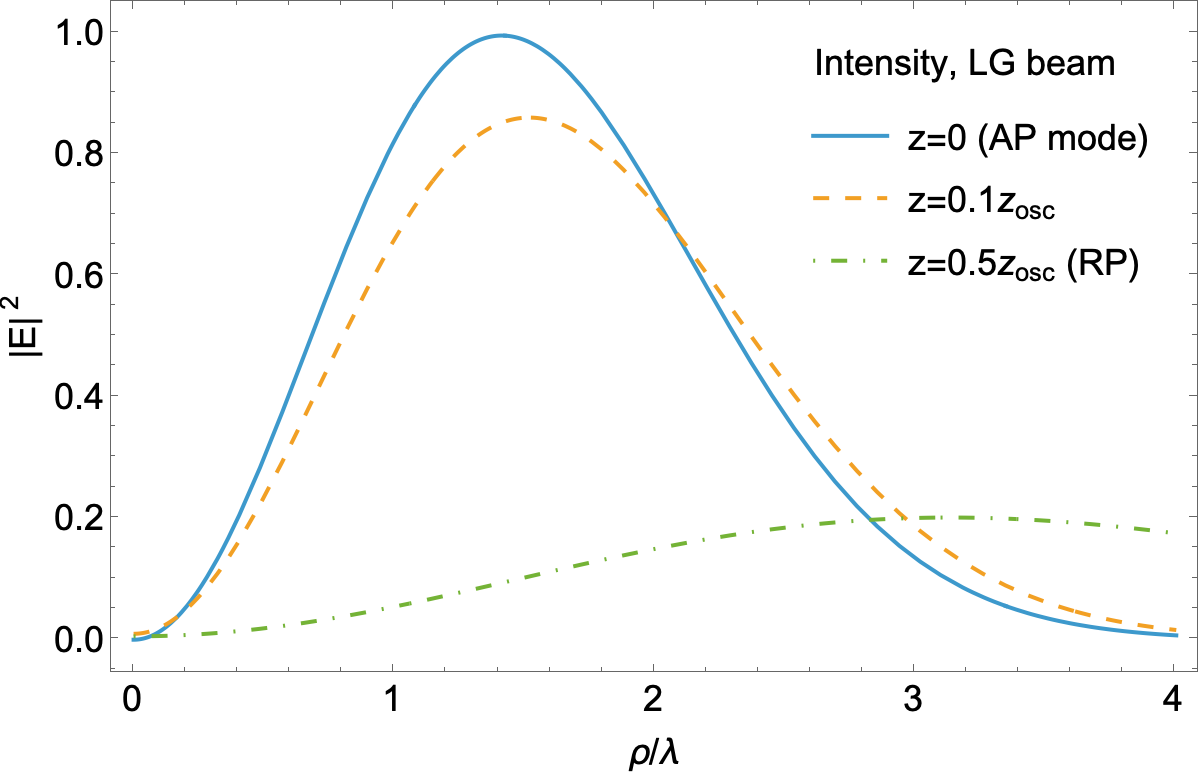}
        \caption{}
    %    \label{fig:b}
    \end{subfigure}
       \centering
       \centering
    \begin{subfigure}[b]{0.45\textwidth}
        \includegraphics[width=\textwidth]{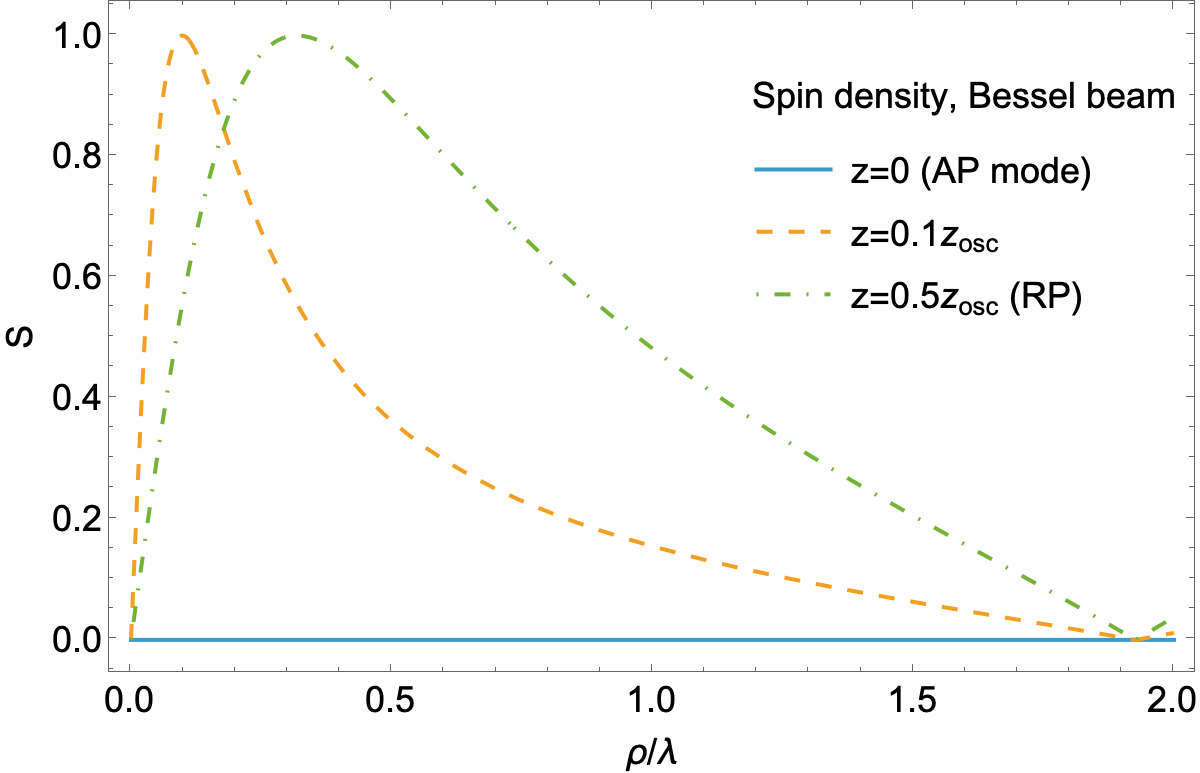}
        \caption{}
        %\label{fig:a}
    \end{subfigure}
    \hfill
    \begin{subfigure}[b]{0.45\textwidth}
        \includegraphics[width=\textwidth]{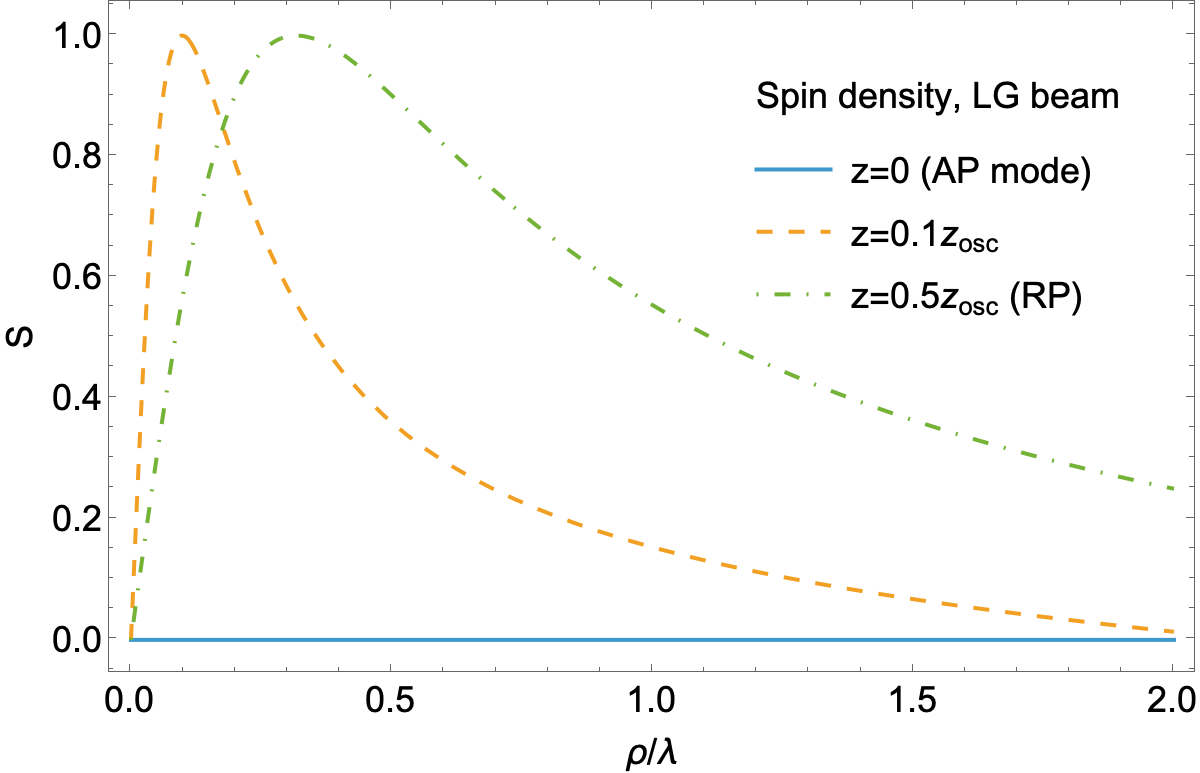}
        \caption{}
        %\label{fig:b}
    \end{subfigure}
\begin{subfigure}[b]{0.45\textwidth}
        \includegraphics[width=\textwidth]{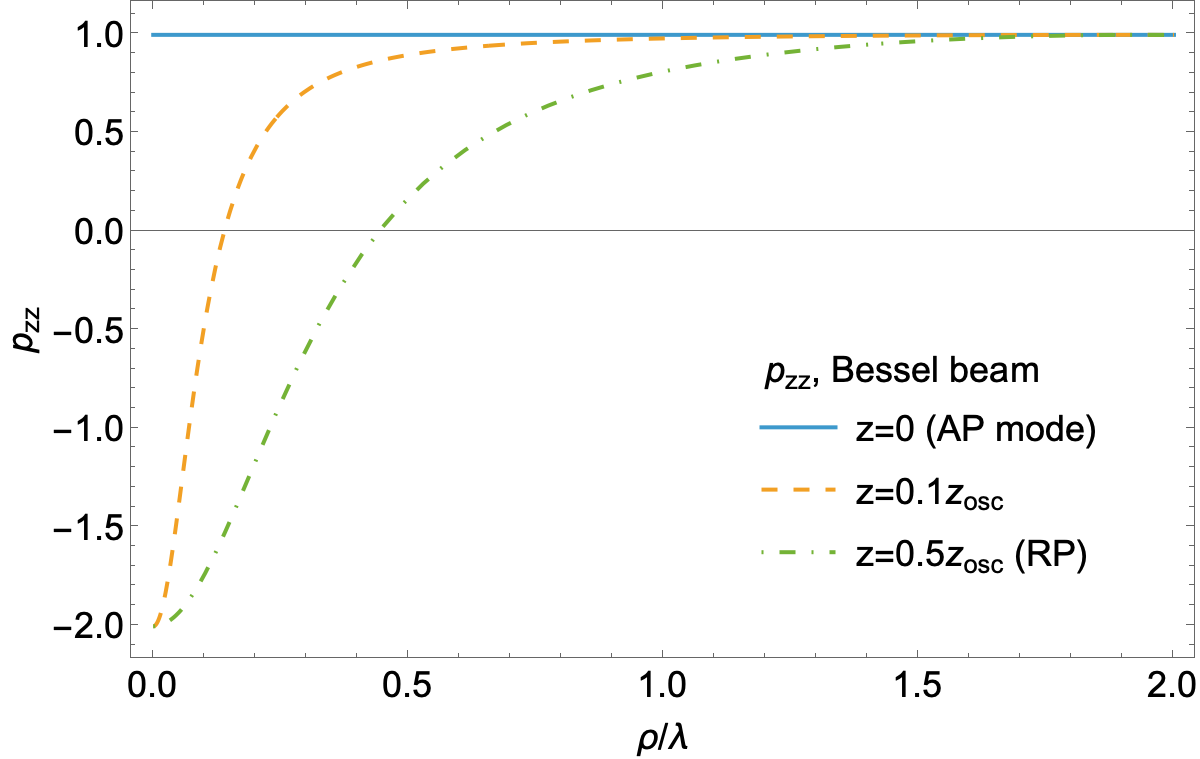}
        \caption{}
        %\label{fig:a}
    \end{subfigure}
    \hfill
    \begin{subfigure}[b]{0.45\textwidth}
        \includegraphics[width=\textwidth]{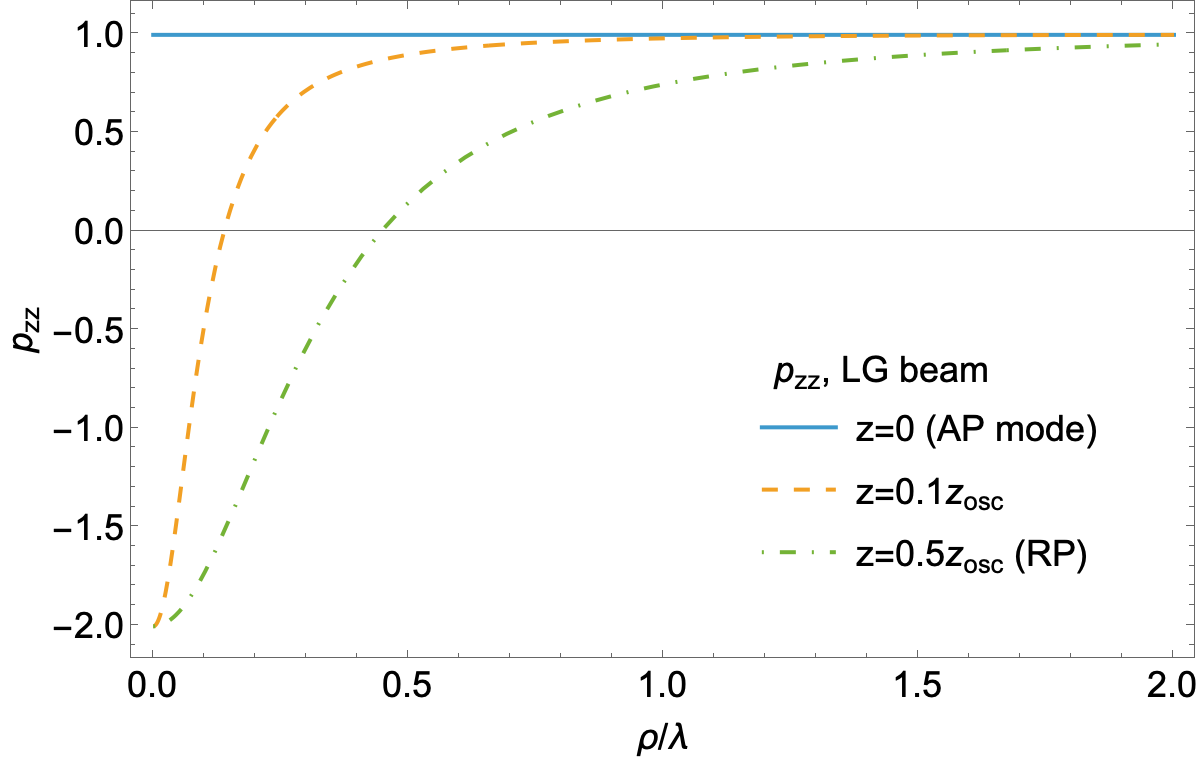}
        \caption{}
        %\label{fig:b}
    \end{subfigure}
    \caption{\raggedright  Dependence of beam polarisation parameters on the propagation distance $z$ for (left column) Bessel and (right column) LG beams. (a,b) Top row: Transverse profile of electric  energy density of the AP mode (at $z=0$) with the distance measured in units of oscillation period $z_{\text{osc}}$. An AP beam at $z=0$ evolves into an RP beam at $z=0.5z_{\text{osc}}$ and converts back to the AP mode at $z=z_{\text{osc}}$.  We choose the Bessel-beam pitch angle parameter $\theta_k=0.2$~rad and $w_0=2\lambda$ for the LG beam waist. (c,d) Middle row: Spin density $S$.
    (e,f) Bottom row:  Alignment parameter $p_{zz}$. }
    \label{fig:edens}
\end{figure}

In the paraxial limit of propagation in an optically active medium (Eq.~(\ref{eq:normmode})), the expressions for spin polarisation simplify. The spin polarisation density depends both on the radial position and the propagation distance, but not on the beam intensity profile:  
\begin{align}
\label{eq:SdenEv}
    |\vec S|^{(AP)}(z)&=\frac{4\bar  k\rho|\sin(z\delta k)|}{4\sin^2(z\delta k)+\bar k^2\rho^2} \, , \\ \nonumber
 |\vec S|^{(RP)}(z)&=\frac{4\bar k\rho|\cos(z\delta k)|}{4\cos^2(z\delta k)+\bar k^2\rho^2} \, ,
\end{align}
and the spin polarisation in the cylindrical basis $(\rho,\varphi,z)$ is  
\begin{align}
\label{eq:SvecEv}
    \vec S^{(AP)}(z)&=\frac{4 \bar k\rho \sin(z\delta k)}{4\sin^2(z\delta k)+\bar k^2\rho^2}(\cos(z\delta k),\sin(z\delta k),0), \\ \nonumber
 \vec S^{(RP)}(z)&=\frac{4\bar k\rho \cos(z\delta k)}{4\cos^2(z\delta k)+\bar k^2\rho^2}(-\sin(z\delta k),\cos(z\delta k),0),
\end{align}
indicating spin-vector rotation during propagation (Fig.~\ref{fig:3DRot} and \ref{fig:edens}c,d).

 It should be noted that the Gauss's law in a paraxial limit, $\vec \nabla_\perp \vec E_\perp(\vec r)+ikE_z(\vec r)=0$ \cite{lax1975maxwell}, locks the relative magnitude and phase between the radial and longitudinal fields, leading to a positive-only azimuthal spin $\varphi$-component, but with no restrictions on a sign of the radial spin that is due to admixture of the AP mode. As a result, a $\varphi$-component of the spin vector remains positive under propagation, while the $\rho$-component has alternating signs.  Spin reaches its maximum value $S=1$  for $\rho=2 |\cos{(z\delta k)}|/\bar k$ for propagating RP beams and $\rho=2 |\sin{(z\delta k)}|/\bar k$ for AP beams, respectively (Fig.~\ref{fig:edens}c,d). This results in the formation of an axially-symmetric C-surface of polarisation singularity ($i.e.$, a surface within a beam where the state of polarisation is 100\% circular) defined by these radial positions (Fig.~\ref{fig:3DRot}). The transverse spatial profile of the spin density is not affected by diffractive expansion of the beam \cite{afanasev2023nondiffractive}. It is interesting that the radius of the C-surface is inversely proportional to the wave vector, therefore, choosing an epsilon-near-zero medium with a small real part of the refractive index \cite{Liberal2017ENZ}, it would be possible to generate non-diffractive spin textures with large transverse sizes.

Since absorption is neglected in our analysis, the spin vector field for the mixed AP-RP modes remains transverse and forms a spin vortex. The circulation of the spin field $\vec S$ can be calculated analytically (Appendix B), and it reaches the maximum value for a pure RP mode in the limit of large radii of the loop $\Gamma \gg \lambda$: $\oint_\Gamma \vec S^{(RP)}d \vec \Gamma =4\lambda $.

 For the alignment parameter, we obtain in the paraxial limit:
\begin{align}
\label{eq:pzzEv}
    p^{(AP)}_{zz}(z)&=\frac{-8\sin^2(z\delta k)+\bar k^2\rho^2}{4\sin^2(z\delta k)+\bar k^2\rho^2}, \\ \nonumber
 p^{(RP)}_{zz}(z)&=\frac{-8\cos^2(z\delta k)+\bar k^2\rho^2}{4\cos^2(z\delta k)+\bar k^2\rho^2}. 
\end{align}
The alignment parameter $p_{zz}$ exhibits periodic oscillations with the propagation distance, while remaining independent of the intensity profile of a beam, thus describing periodic transformations from RP to AP modes and back with related disappearance and re-appearance of longitudinal field $E_z$ (Fig.~\ref{fig:edens}e,f). The longitudinal field component $E_z$ is dominant in the beam centre for pure RP and mixed RP-AP modes, but it is absent for AP beams, as expected. The range of the radii where $E_z$ is same or larger in magnitude than the transverse field $E_\perp$ varies between $\rho=\lambda/\pi$ for RP and $\rho=0$ for AP mode, respectively. 
Such re-distribution of energy density across the beam profile due to the varying polarisation is a manifestation of spin-orbit coupling in the optical fields \cite{bliokh2015spin}. 
%Such re-distribution of energy density across the beam profile due to the varying polarisation may be viewed as generalised spin Hall effect of light \cite{bliokh2010angular, bliokh2015quantum}. 

\subsection{Diffractive Laguerre-Gauss CVBs}

In order to demonstrate independence of  the transverse spin dynamics on diffraction during propagation, we now consider paraxial monochromatic Laguerre-Gauss (LG) beams. The beam propagates in the $z$-direction with a topological charge $l$ and zero radial index ~\cite{saleh2008fundamentals,capolino2024}.  The electric field components of the LG beam in the transverse ($\rho,\varphi$)-plane in cylindrical coordinates are given by
\begin{equation}
\label{eq:Eperp}
\vec{E}_\perp= \vec{\eta}_\perp A_\perp \frac{w_0}{w(z)} \bigg(\frac{\rho}{w(z)}\bigg)^{|l|}  e^{-\frac{\rho^2}{w^2(z)}} e^{i(l\varphi + kz + k \frac{\rho^2}{2R(z)} -(l+1)\psi(z)-\omega t)} , 
\end{equation}
where, the vector $\vec{\eta}_\perp= \hat{\vec\rho} (\hat{\vec \varphi})$ defines the RP(AP) polarisation of the beam in the ($\rho,\varphi$)-plane, $A_\perp$ is a normalisation constant, $w_0$ is the beam waist at $z=0$,
$R(z)$ is the beam curvature radius, $\psi(z)$ is the Gouy phase factor, and $w(z)=w_0\sqrt{1+\big(\frac{z}{z_R}\big)^2}$, where $z_R=kw_0^2/2$ is the Rayleigh length. 
The longitudinal component $E_z$ is obtained  from the Gauss's law in a paraxial limit \cite{lax1975maxwell}:
$\vec \nabla_\perp \vec E_\perp(\vec r)+ikE_z(\vec r)=0$.

Normal modes for LG beams can be identified as the states with a definite OAM ($l$) and circular polarisation $(\sigma)$, i.e., $|l=\mp1,\sigma=\pm1>$. We use these states as building blocks to construct AP and RP modes as in Eq.~(\ref{eq:APRP0}) and analyse their evolution described by Eq.~(\ref{eq:APev}).   For a special case of a central region of the propagating beam,  $\rho<w$, we use a Taylor expansion and recover the similar field expressions as in Eqs.~(\ref{eq:EAPpar}) and (\ref{eq:ERPpar}) for the Bessel beam, up to an overall polarisation-independent factor. The same evolution therefore is observed for the considered polarisation parameters in the LG beam central region ($\rho<w$) as obtained in the paraxial limit for the AP and RP Bessel beams in Eqs.~(\ref{eq:pzzEv}--\ref{eq:SdenEv}).
A beam waist of the LG beam expands with the propagation due to diffraction (Fig.~\ref{fig:3DRot}b), whereas the C-surface of polarisation singularity ($S=1$) for both LG and Bessel beams oscillates between the radial positions $0\leq \rho\leq\lambda/\pi$, synchronised with the periodic conversion between AP and RP modes (Fig.~\ref{fig:edens}e,f). The spin rotates in the transverse plane, reversing sign when the beam passes through a pure AP mode 
%(see animations of the transverse spin and spin density during the beam propagation in Supplementary Materials), 
while its radial component changes sign when passing through a pure RP mode. Despite diffraction in LG beams, the spin behaviour and longitudinal-transverse field interplay remains the same as for non-diffractive Bessel beams.

While Bessel beams undergo AP–RP mode oscillations without diffraction, LG beams exhibit both the mode conversion and transverse broadening.  Similarly, 
the evolution of the alignment parameter $p_{zz}$, quantifying the relative weight of longitudinal and transverse electric field components, also oscillates during AP–RP transformations but does not experience diffractive spreading (cf. $p_{zz}$ for Bessel and LG beams; non-diffractive nature of this parameter in free space was shown in Ref.~\cite{afanasev2023nondiffractive}). The longitudinal field contribution is maximal at the beam centre ($\rho=0$) and in the RP mode ($z=0.5z_\text{osc}$).

While the spin density vanishes in the pure AP mode, any mixed RP–AP state supports a polarisation singularity (C-line), corresponding to a transverse cross-section of the C-surface (Fig.~\ref{fig:edens}e,f). The C-line forms a circle whose radius varies from zero in the AP mode to a maximum of
$\lambda/\pi$ in the RP mode. For $\rho < \lambda$, both polarisation parameters show nearly identical behaviour for LG and Bessel beams, and the validity range of the paraxial approximation in Eqs.~(\ref{eq:ERPpar},\ref{eq:EAPpar}) increases with propagation for LG beams.

\begin{figure}[th]
    \begin{subfigure}[b]{0.345\textwidth}
        \includegraphics[width=\textwidth]{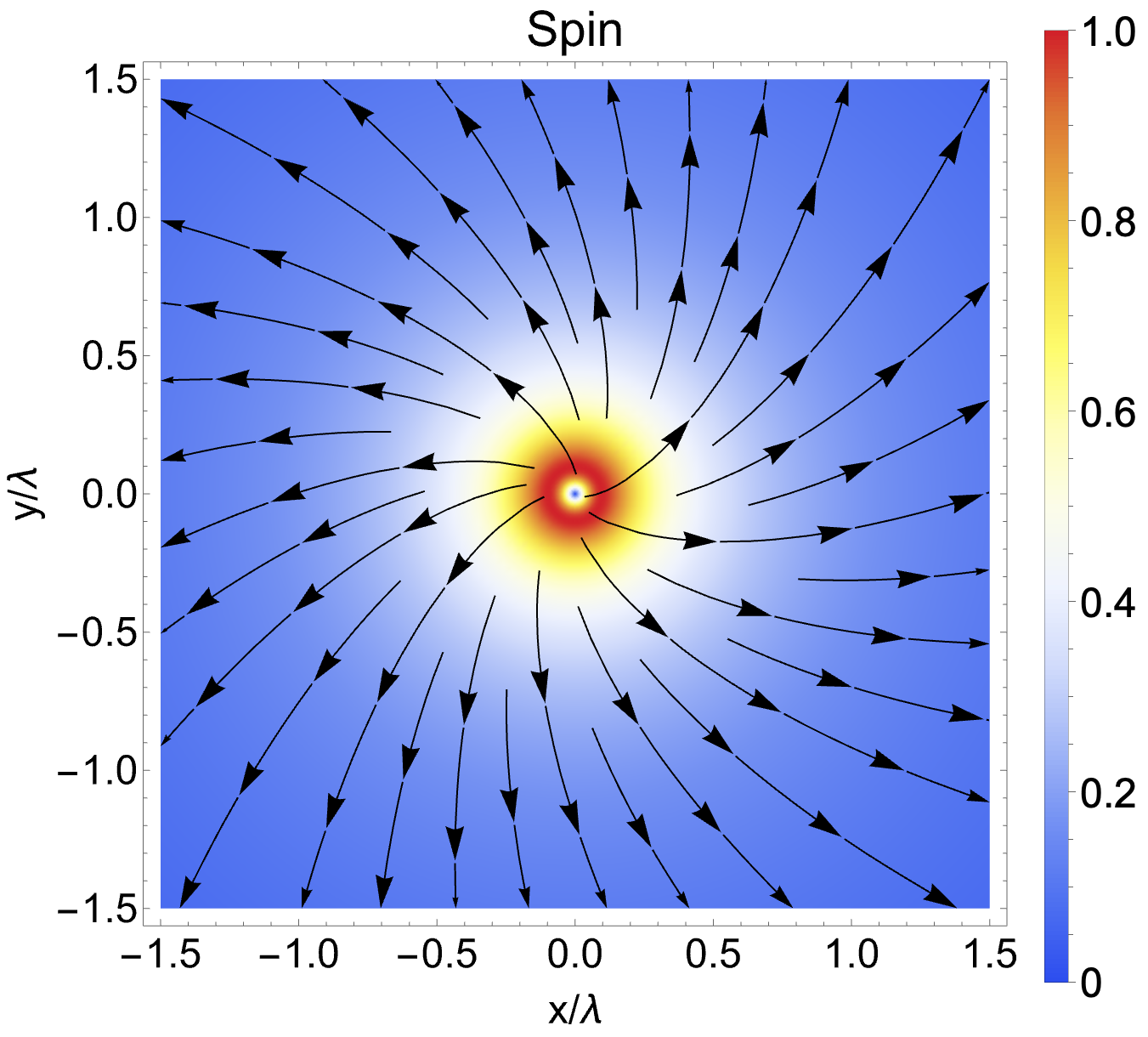}
        \caption{}
        \label{fig:a}
    \end{subfigure}
    \hfill
    \begin{subfigure}[b]{0.31\textwidth}
        \includegraphics[width=\textwidth]{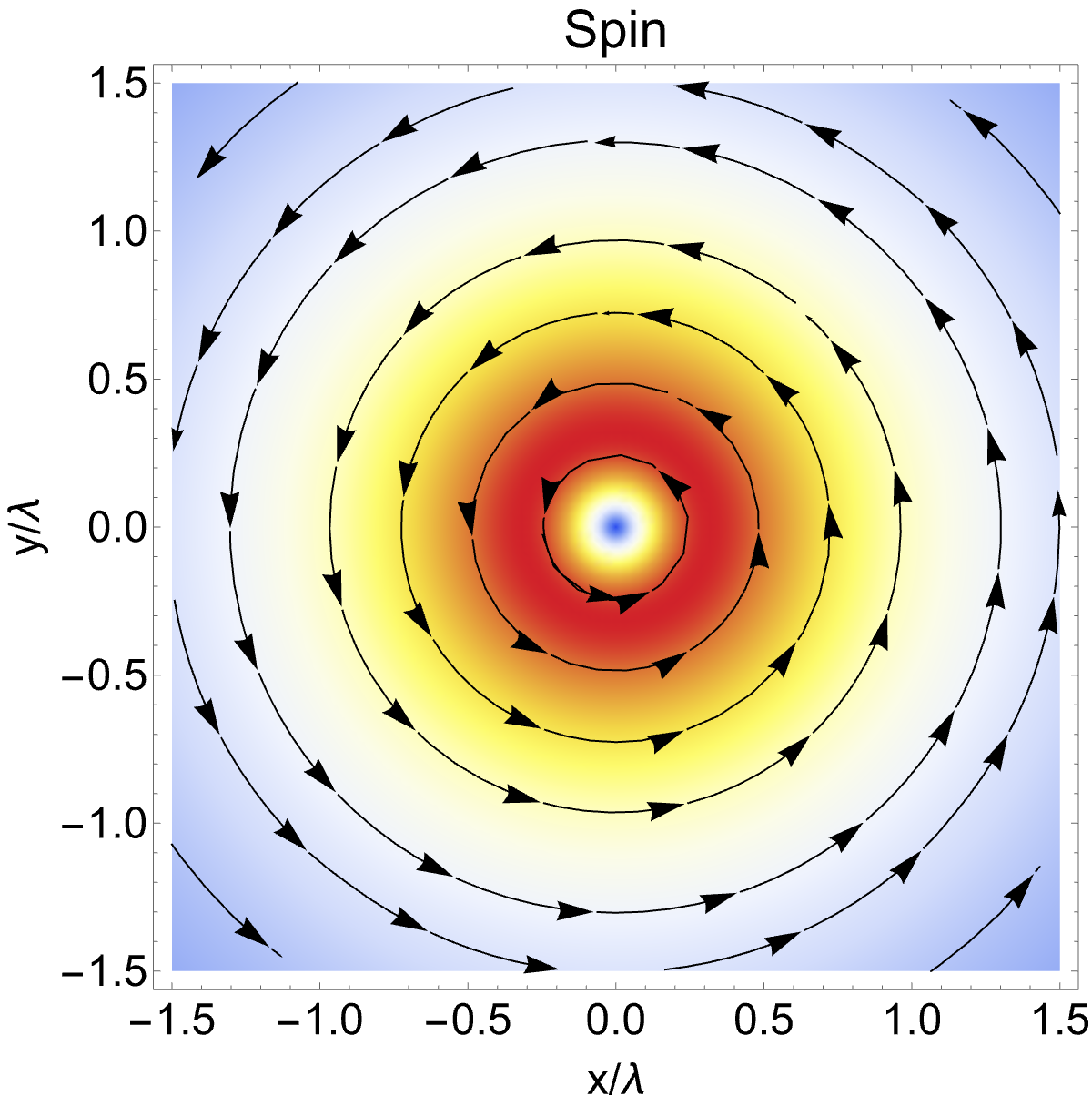}
        \caption{}
        \label{fig:b}
    \end{subfigure}
    \hfill
        \begin{subfigure}[b]{0.31\textwidth}
        \includegraphics[width=\textwidth]{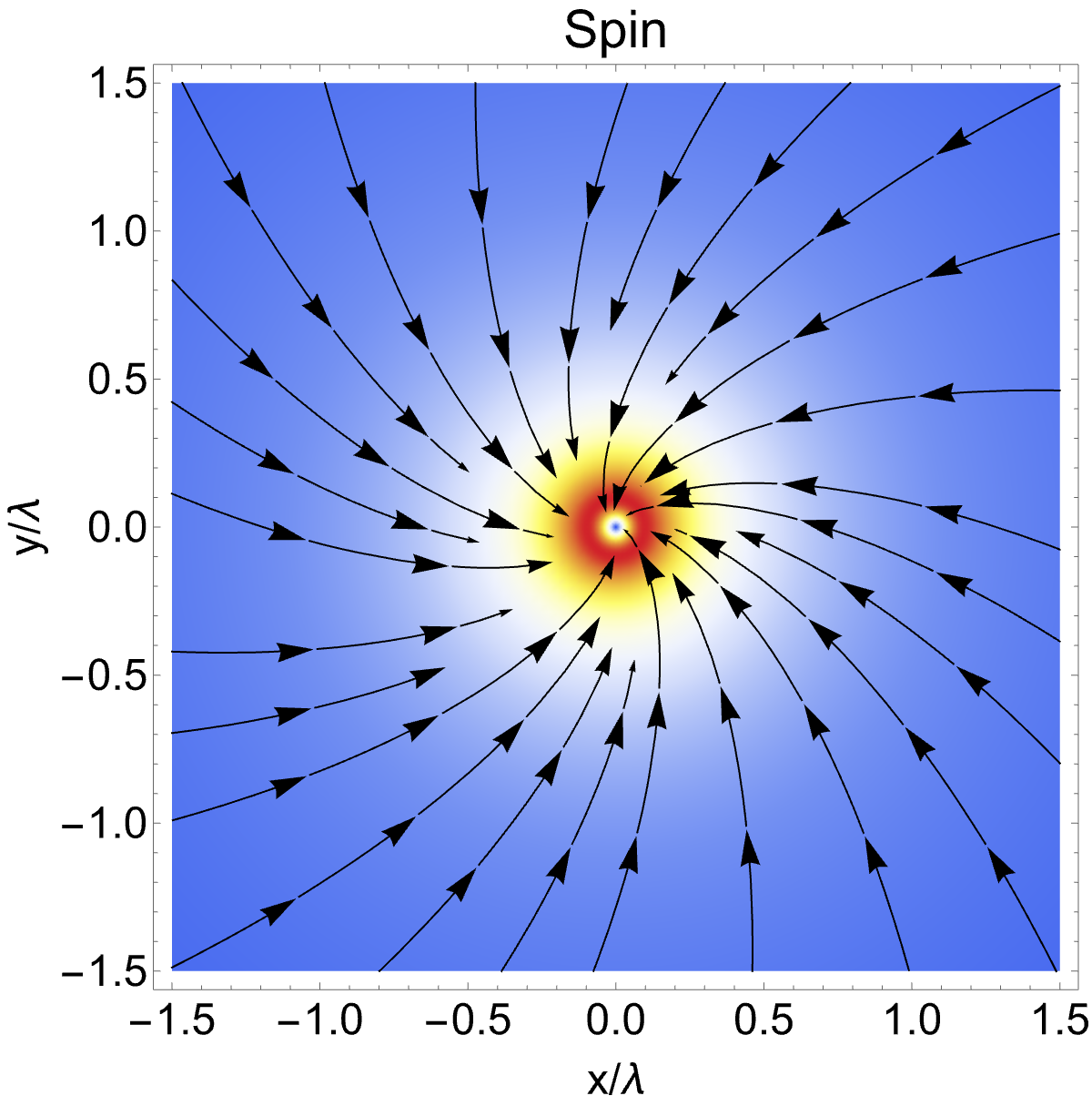}
        \caption{}
        \label{fig:a}
    \end{subfigure}
    \hfill
    \caption{Spin density $S$ (color coded) and spin orientation (arrows) across the wave front for an AP beam after the propagation distances of $z=0.1z_{\text{osc}}$ (a), $z=0.5z_{\text{osc}}$ (b) and $z=0.9z_{\text{osc}}$ (c). The spin vector is in the transverse plane and oriented along a spiral for mixed RP-AP modes and along a azimuthal (radial) direction for a pure RP (AP) mode, with a radial component flipping a sign when the beam passes through either a RP or an AP pure mode. \raggedright The calculations are for a levorotary material. }
      \label{fig:spin2D}
\end{figure}

Figure~\ref{fig:spin2D} demonstrates evolution of the optical spin. The spin vector lies in the transverse ($\rho,\varphi$)-plane and rotates clockwise or anti-clockwise depending on the sign of the rotary power, similar to the rotation of a polarisation plane of linearly polarised light for dextrorotary or levorotary chiral materials \cite{saleh2008fundamentals}. A major difference for the spin is the existence of a ``forbidden" region of negative azimuthal components. The spin rotates continuously between 0 and 180$^{\circ}$ with respect to the radial direction, then instantaneously flips the sign and continues to rotate in the same direction, by-passing the negative region of $S_\varphi$ components. In contrast, linear polarisation for plane waves rotates continuously under optical activity.

We shall draw a comparison between the transverse spin rotation of light and parity-violating rotation of transverse neutron spin in the scattering off nuclei due to admixture of electroweak interactions \cite{forte1980first,sushkov1982parity,klein1983neutron}. Neutrons are spin-1/2 particles, and their transverse spin rotation is caused by the difference of cross sections of neutrons with opposite helicities $\pm1/2$ scattering on unpolarised or spin-zero nuclei. A similar spin effect is not possible for the plane-wave light--with linear polarisation plane rotation being the closest analogue--but this possibility is realised for the vectorial vortex light outlined in this study.

%Note that the tip of spin vector only sweeps first two quadrants in ($\rho,\varphi$)-plane, with a $\varphi$-component remaining positive during propagation.

%Evolution of polarisation of the resulting field as a function of propagation distance is presented in the figures below starting with  azimuthal polarisation at $z$=0.

\subsection{Numerical analysis}
\begin{figure}[tbh]
    \includegraphics[width=1\linewidth]{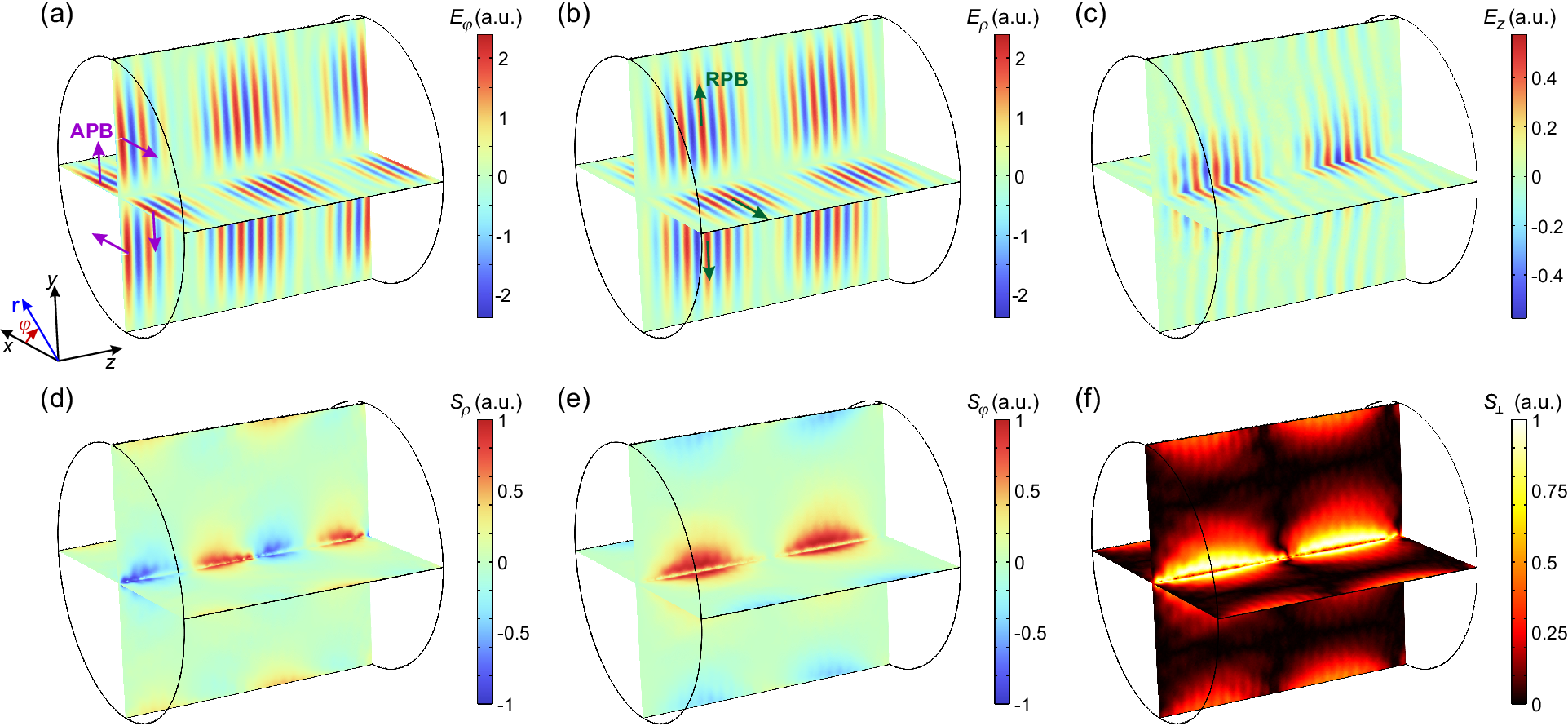}
    \caption{\raggedright Numerically simulated field maps of azimuthal $E_\varphi$ (a), radial $E_\rho$ (b) and longitudinal $E_z$ (c) electric field components,  radial $S_\rho$ (d),  azimuthal $S_\varphi$ (e) and absolute value of total transverse $S_\perp$ (f) spin components for an AP beam at a wavelength of 650 nm propagating in a 7.5-$\upmu$m-thick layer of an optically active medium.}
    \label{fig:num}
\end{figure}

Numerical simulations of interaction of cylindrical beams with an optically active medium were performed using a finite element method (COMSOL Multiphysics software). A weakly diverging ($w_0=10\lambda /\pi$) azimuthally-polarised LG beam having a wavelength of 650~nm and described by a formalism from Ref.~\cite{capolino2024} was focused in the middle of a chiral layer. The constitutive relations in the chiral medium were modified to be (within the $i \omega t$ notations adopted in the software):
\begin{align}
    \vec D&=\varepsilon \vec E - i \frac{\xi}{c} \vec H,\\ \nonumber
    \vec B&=\mu \vec H + i \frac{\xi}{c} \vec E,
\end{align}
where for simplicity $\varepsilon$ was taken to be 1 and $\xi$ was adjusted in such a way that two full 360$^{\circ}$ polarisation rotations of a plane wave would take place across the modeled chiral layer having a thickness of 7.5~$\upmu$m, which was determined to be the maximum permitted by the computational complexity of the simulations. The rotary power of the optically active medium in this case is twice larger than the one used in the analytical calculations, as the optical activity parameter $\xi$ was also incorporated into the constitutive relation for the magnetic field. The simulation domain was chosen to be cylindrical to match the cylindrical symmetry of the beam. To avoid back reflections it was surrounded by perfectly matching layers (PMLs) from all the sides, apart from the front boundary acting as a source. Analysing the numerically calculated field distribution, it was specifically checked that the reflection from the chiral layer is negligible, which is expected as the contrast of the optical properties between the layer and the surroundings, entirely determined by $\xi$, is very small. The results of the numerical simulations are presented in Fig.~\ref{fig:num}. Four full radial-to-azimuthal inter-conversions of the beam across the optically active layer can be seen analysing the azimuthal and radial components of the fields (Fig.~\ref{fig:num}a,b). Simultaneously, a longitudinal component, not present in the incident azimuthally-polarised beam, appears in the regions when the latter is partially or fully converted to the radial counterpart (Fig.~\ref{fig:num}c). The maps of the spin components of the fields presented in Fig.~\ref{fig:num}d--f were found to be in excellent agreement with the theoretical predictions.

\section{Experimental observation of radial-azimuthal beam oscillations}
\begin{figure}[h]
    \includegraphics[width=1\linewidth]{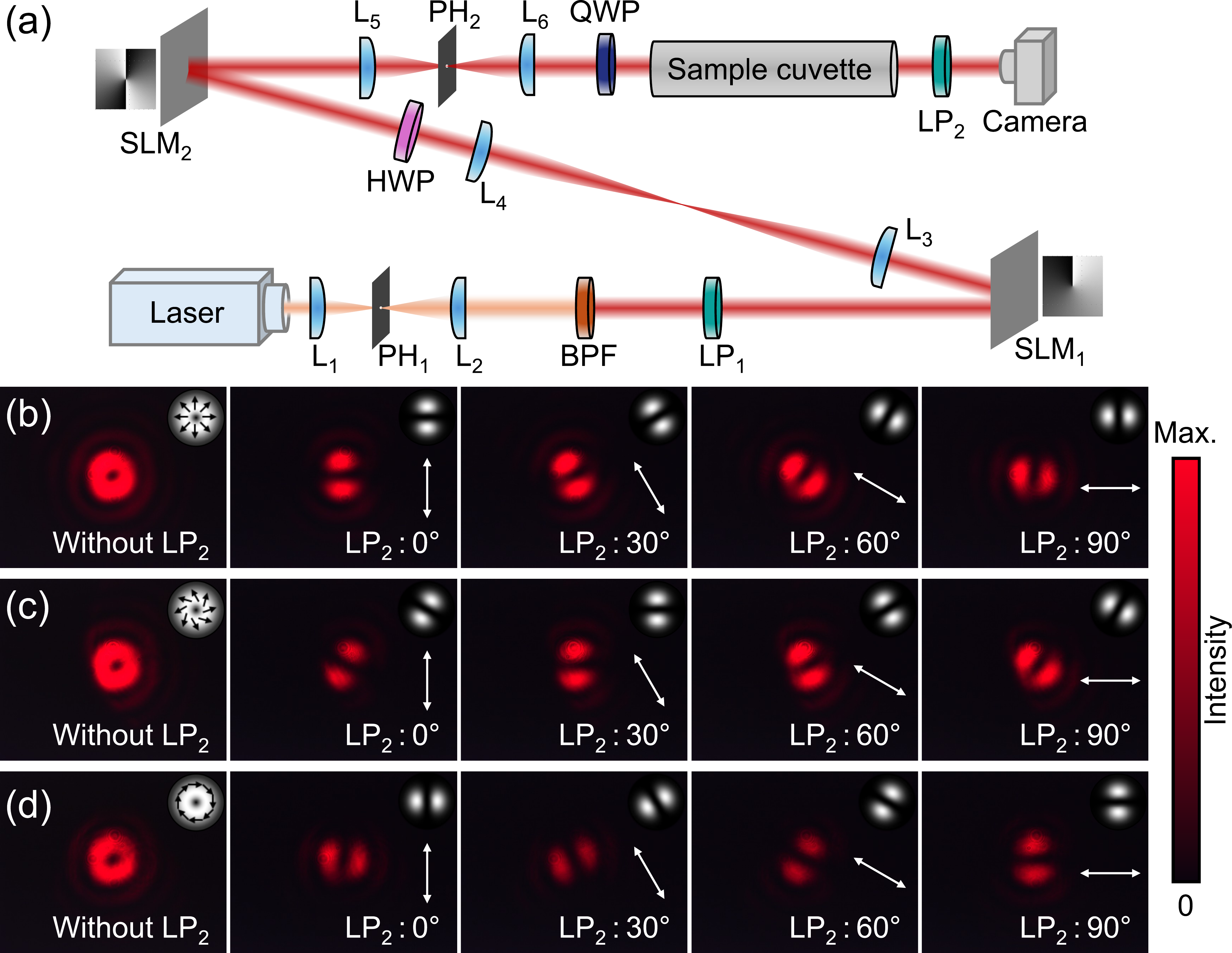}
    \caption{\raggedright (a) Schematic of the optical setup for measuring cylindrical beam inter-conversion, comprising a supercontinuum white light laser, lenses (L), pinholes (PH), linear polarisers (LP), spatial light modulators (SLM), a half-wave plate (HWP), a quarter-wave plate (QWP), a bandpass filter (BPF) and a cuvette with an optically active medium. (b--d) Transverse beam profiles recorded for different analyser orientations (b) in the absence of optically active medium, and for (c) 5~cm and (d) 15~cm beam paths through an optically-active medium based on high-concentration D-fructose solution. Inserts in the left column illustrate the beam intensity profile with arrows indicating the direction of the electric field; in other columns the inserts show the orientation of the analyser LP$_2$ and the beam images expected for this orientation.}
    \label{fig:exp}
\end{figure}

To experimentally verify the inter-conversion between the AP and RP beams during their propagation in an optically active medium, we measured the polarisation state of CVBs at a wavelength of 650~nm propagating through a high-concentration  (0.727~\text{g/mL}) D-fructose solution. The beam path within the medium was varied using cuvettes of different lengths (Fig.~\ref{fig:exp}a). A RP beam was produced from a laser beam (Supercontinuum Fianium Femtopower 1060 SC450-2, spectrally filtered to a central wavelength of 650~nm with a bandwidth of 40~nm) by sequentially passing it through a linear polariser ($\text{LP}_1$) set at $0^\circ$, a half-wave plate set at $\pi/8$, and two spatial light modulators (PLUTO2-NIR011, HOLOEYE) for generation of topological charges of $l_1=1$ and $l_2=-2$, respectively, followed by a quarter-wave plate oriented at $\pi/4$ \cite{aita2025longitudinal}. The polarisation state of the beam that passed through the cuvette was analysed using a linear polariser, while the resulting intensity profiles were imaged using a CCD camera. 

The polarisation analysis of the incident beam shows a pure RP state with an intensity pattern that has two lobes symmetric with respect to the beam axis following the orientation of the analyser axis, which corresponds to the electric field oscillation in the radial direction (Fig.~\ref{fig:exp}b). For a beam propagating in the D-fructose solution over a 5~cm, the optical activity of the medium induces a $30^\circ$ local polarisation rotation, resulting in a mixed AP-RP beam polarisation state, with the intensity pattern not aligned but rotated by $30^\circ$ with respect to the analyser orientation (Fig.~\ref{fig:exp}c). By increasing the path length to 15~cm, a full conversion of the incident RP beam to AP beam is achieved: the intensity map shows lobes oriented in the direction perpendicular to the analyser axis, which marks the azimuthal direction of the electric field (Fig.~\ref{fig:exp}d). Overall, the observed mode conversion corresponds to $z_{\text{osc}}$= 30~cm.

\section{Summary and Discussion}
We analysed theoretically, numerically and experimentally the evolution of polarisation of CVBs propagating in isotropic optically active media. Propagation normal modes were identified and shown to have different propagation constants, leading to predictions for inter-conversion between AP and RP modes with a spatial period defined by the same rotary power of the medium as for plane waves. 

For both Bessel and LG beams, we observed the ``pulsating" spin density in the propagating beam, which reached a circular polarisation singularity ($S=1$, or C-surface) at the $\rho=\lambda/\pi$ radial position for the RP modes, with the C-line shrinking to a zero radius for the AP modes. Spin orientation changes between purely azimuthal for the RP mode and radial for the AP mode, while its azimuthal component remains always {\it positive}. The latter observation means that the tip of the spin vector periodically swipes only a half-circle in the ($\rho,\varphi$)-plane, being constrained by Gauss's theorem that locks a relative phase between the longitudinal and the transverse electric fields. The direction of the spin rotation matches the clock-wise (counter-clockwise) direction of polarisation plane rotation for plane waves in the same dextrorotary (levorotary) material. A spin-flip effect for the beam passing through the pure AP mode may be exploited to enhance sensitivity of probing the chirality of matter using CVBs.

The developed approach is extended to electromagnetic multipole radiation that shows vortex-like spin structures near field-zero regions \cite{neugebauer2019emission,vernon2024non}. It implies, for example, that inter-modal oscillations between electric-dipole (E1) and magnetic-dipole (M1) radiation would take place in optically active media and the transverse spin would rotate similar to the case of the cylindrical vector modes (see Appendix C for details).

The presence of transverse optical spin in CVBs does not necessarily imply that the transverse spin would rotate under optical activity. Indeed, inspecting Eq.~(\ref{eq:normmode}), one can see that a normal mode does contain a transverse spin component, but the mode propagates without changing its state of polarisation. Evolution of transverse spin takes place in superposition of normal modes due to their different propagation constants. Another anticipated effect will be circular dichroism that would result in a higher rate of absorption of one of the normal modes with respect to another and, since the normal modes have opposite spins, in development of a longitudinal spin component during CVB propagation. The polarisation normal modes can be also associated with optical spin merons \cite{annenkova2025universal} that are invariant under propagation. Rotation of photon transverse spin in optically active media resembles a similar effect of neutron spin rotation induced by parity-violating weak interactions  \cite{forte1980first,sushkov1982parity,klein1983neutron}, emphasising particle-like behavior of the photon spin.

\section*{Acknowledgments}

Y.X., A.V.K. and A.V.Z. acknowledge support from the UK EPSRC projects EP/W017075/1 and UKRI3056. A.A. was supported by the US ARO Award W911NF-23-1-0085.

\bibliography{vortex}

\section*{Appendix A. Identifying propagation normal modes}

Here, we provide details of identifying the normal modes for propagation of structured electromagnetic waves in an optically active medium. 
%The solution for the plane waves is known: polarisation state of left- and right circular polarisation (LHC and RHC). The approach is as follows: (a)~recognize (and verify) that circularly polarised Bessel beams are normal modes for propagation in optically active media, (b)~expand azimuthal and radial modes in terms of LHC and RHC Bessel solutions, and (c)~analyse the resulting evolution of polarisation under propagation.
The effect of optical activity arises  in the first-order expansion of nonlocal material relation between $\vec D$ and $\vec E$ fields as a result of spatial dispersion \cite{landau2013electrodynamics}. In an isotropic optically active medium, the electric displacement is 
\begin{equation}
    \vec D = \epsilon \vec E+ \epsilon_0\xi \vec \nabla\times\vec E,
    \label{eq:consitutive}
\end{equation}
where $\epsilon$ is the permittivity of a medium and $\xi$ is the pseudoscalar quantity which describes optical activity.
For the plane-wave solution of Helmholtz equation, $\vec E=\vec E_0 \exp[i (\vec k \vec r -\omega t)]$, we replace $\vec\nabla\times\vec E\to i\vec k\times\vec E$ and diagonalise the material equation Eq.~(\ref{eq:consitutive}) to obtain  propagation normal modes
\begin{equation}
   \vec E_{\pm}=E_0\vec e_{\pm} \exp[i (\vec k_{\pm} \vec r -\omega t), 
   \label{eq:PW}
\end{equation}
where unit vectors $\vec e_{\pm}=-({\pm \hat {\vec x}}+ i \hat{\vec y})/\sqrt2$ correspond to RCP and LCP polarisation, respectively, and the propagation constants  $k_\pm = n_\pm \omega/c$ contain refractive indices of the normal modes
\begin{equation}
 n_\pm=\sqrt{n^2\pm G}   
\end{equation}
with $n^2=\epsilon/\epsilon_0$ and $G=\xi  k$ (see, for example, Ref.~\cite{saleh2008fundamentals}). It follows that
\begin{equation}
 \vec D=\epsilon_0 n_\pm ^2 \vec E \, .
 \label{eq:prop}
\end{equation}
In order for the linear relation Eq.~(\ref{eq:prop}) to hold, it is essential that the electric field of the normal mode is an eigenstate of the curl operator $\vec\nabla\times$ :
\begin{equation}
\label{eq:nmodeAp}
 \vec\nabla\times\vec E_\pm =\pm k_\pm \vec E_\pm \, , 
\end{equation}
and we will use this relation as a criterion to identify propagating normal modes of the polarised structured waves.

\section*{Appendix B. Circulation of spin in a propagating CVB}

We calculate circulation of the spin vector $\vec S$ around the beam axis along a closed circular contour $\Gamma$ of radius $\rho_0$ using Eq. (\ref{eq:SvecEv}) for the spin vector:
\begin{align}
\oint_\Gamma \vec S^{(AP)}d \vec \Gamma =
\begin{cases}
        0, & \rho_0\to 0\\
      2\lambda|\sin{(z\delta k)}|, & \rho_0= \lambda/\pi,\\
4\lambda|\sin{(z\delta k)}|, & \rho_0\gg \lambda \\
\end{cases}
\end{align}    

for an AP beam and

\begin{align}
\oint_\Gamma \vec S^{(RP)}d \vec \Gamma =
\begin{cases}
        0, & \rho_0\to 0,\\
      2\lambda|\cos{(z\delta k)}|, & \rho_0= \lambda/\pi,\\
4\lambda|\cos{(z\delta k)}|, & \rho_0\gg \lambda\\
\end{cases}
\end{align}   
for a RP beam. It follows that during propagation the spin circulation reaches maximum when the beam oscillates into a pure RP mode with the spin circulation equal to $2\lambda$ on the C-line and $4\lambda$ for large contours $\rho_0 \gg \lambda$. The circulation is zero for a pure AP mode. The sign of spin circulation remains positive for mixed RP-AP modes.

\section*{Appendix C. Dipole radiation in an optically active medium}

%-Propagation normal modes result from superposition of radiation from electric and magnetic dipoles;

%-Radiation from electric or magnetic dipoles is a linear superposition of normal modes; near the minimum, its polarisation is similar to the radially or azimuthally polarised cylindrical beams, respectively;

%-A formula+figure demonstrate evolution of the transverse spin of electric/magnetic dipole radiation in optically active media

An oscillating point-like electric dipole with a dipole moment $\vec p$ produces a radiated field described by~\cite{Jackson1999}
\begin{equation}
    \vec{E}_{\text{E1}}=\frac{1}{4\pi\epsilon_0}\bigl\{k^2 (\vec n\times\vec p)\times\vec n\frac{e^{ikr}}{r}+[3\vec n(\vec n\vec p)-\vec p]\bigl(\frac{1}{r^3}-\frac{ik}{r^2})e^{ikr} \bigr\} \, ,
\end{equation}
where $\vec n=\vec r/r$. The transverse spin of electric dipole radiation develops around the dipole axis  \cite{neugebauer2019emission, vernon2024non} and its morphology is similar to RP modes of CVBs.
This field does not satisfy the normal mode condition for propagation in an optically active medium (see Appendix A). However, normal modes can be obtained as a linear combination of this field with a field of an oscillating {\it magnetic} dipole having a magnetic dipole moment $\vec m$ proportional to the electric dipole moment $\vec m=c\vec p$~\cite{kerker1983electromagnetic}: 
\begin{equation}
    \vec E_{\text{M1}}=-\frac{1}{4\pi\epsilon_0}k^2 (\vec n\times\vec p)\frac{e^{ikr}}{r}\bigl(1-\frac{1}{ikr}\bigr).
\end{equation}
The normal modes are duality-symmetric superpositions of these two fields~ \cite{zambrana2013duality}: 
\begin{equation}
    \vec E_{\pm}=-\frac{(\pm\vec E_{\text{E1}}+i\vec{E}_{\text{M1}})}{\sqrt{2}},
\end{equation}
and the electric and magnetic dipole radiation can be expressed in terms of the normal modes as
\begin{equation}
\label{eq:E1M1normal}
 \vec E_{\text{E1}}=\frac{1}{\sqrt{2}}[-\vec E_++\vec E_-], \ \ \
 \vec E_{\text{M1}}=\frac{i}{\sqrt{2}}[ \vec E_++\vec E_-].  
\end{equation}
Since in the optically active medium the normal modes have different propagation constants $k_+\neq k_-$, the E1 field transforms in the M1 counterpart after propagating over the radial distance $r_{osc}=\pi/(k_--k_+)$ and then transforms back in the E1 field after the same propagation distance. Therefore, an observer at a given distance from the dipole source may detect a pure E1 mode, while another observer at a radial distance shifted by a half-oscillation length $r_{osc}/2$ would observe a pure M1 mode.  

In the far-field region and away from the dipole axis, the dipole radiation is linearly polarised, and normal-mode decomposition in Eq.~(\ref{eq:E1M1normal}) predicts familiar rotation of the polarisation plane under the propagation~\cite{saleh2008fundamentals}. In the vicinity of the dipole axis $\vec n$, the spin density of the dipole radiation and the direction of a spin polarisation vector  undergo the oscillations similar to those observed for CVBs described by Eqs.~(\ref{eq:pzzEv})--(\ref{eq:SdenEv}), where the E1 field is associated with the RP mode, and the M1 field with the AP mode.

It should be noted that transverse spin density of electric-dipole radiation was experimentally observed in Ref.~\cite{neugebauer2019emission}, while we are not aware of the experimental studies for dipole spin's propagation in optically active media. 
It is straightforward to extend the described approach to higher electric or magnetic multipoles with more complex polarisation topologies~\cite{vernon2024non}.
%%%%%%%%%%%%%%%%%%%%%%%%%%%%%%%%%%

%\begin{figure}[h]
%    \includegraphics[width=0.32\linewidth]{Figures/pzz2D-z0.png}
%    \includegraphics[width=0.32\linewidth]{Figures/pzz-2D-z01.png}
%    \includegraphics[width=0.32\linewidth]{Figures/pzz2D-z05.png} 
%    \caption{Alignment parameter $p_{zz}$ across the wave front for different propagation distances.}
%    \label{fig:pzz2D}
%\end{figure}

%\begin{figure}[h]
%    \includegraphics[width=0.32\linewidth]{Figures/spin2D-z0.png}
%    \includegraphics[width=0.32\linewidth]{Figures/spin2D-z01.png}
%    \includegraphics[width=0.32\linewidth]{Figures/spin2D-z05.png} 
%    \caption{Spin density across the wave front for different propagation distances. Polarisation reaches its maximum value S=1 (C-circle) at a position $\rho=\lambda/\pi$ for pure radial polarisation; the radius of C-circle reduces to zero in the limit of azimuthal polarisation.}
%    \label{fig:spin2D}
%\end{figure}

%No data were generated or analysed in the presented research.

%\paragraph{Supplemental Information.} See Supplement for supporting content.

%\paragraph{Acknowledgements.}

%%%%%%%%%%%%%%%%%%%%%%%%%%%%%%%%%%

%\bibliographyfullrefs{vortex}

%\newpage

%\input{Supplemental}

\end{document}